\documentclass{ifacconf}

\usepackage{graphicx}      
\usepackage{natbib}        
\usepackage{enumerate}
\usepackage{amsmath}
\usepackage{amssymb}
\usepackage{amsfonts}
\usepackage{float}
\usepackage{xspace}
\usepackage{color,colortbl}
\usepackage{url}
\usepackage{epstopdf}
\usepackage{subfigure}
\usepackage{makecell}
\usepackage{booktabs}
\usepackage{graphicx}
\usepackage{multirow}
\graphicspath{./figures/}

\begin{document}
\begin{frontmatter}

\title{Traffic-Aware Eco-Driving Control in CAVs via Learning-based Terminal Cost Model\thanksref{footnoteinfo}} 

\thanks[footnoteinfo]{This work is supported by the United States Department of Energy, Advanced Research Projects Agency – Energy (ARPA-E) NEXTCAR project (Award Number DE-AR0000794).}

\author[First]{Mehmet Fatih Ozkan} 
\author[First]{Dennis Kibalama} 
\author[Second]{Jacob Paugh}
\author[Second]{Marcello Canova}
\author[Second]{Stephanie Stockar}

\address[First]{Center for Automotive Research, The Ohio State University, 
   Columbus, OH 43212 USA (e-mail: ozkan.25@osu.edu, kibalama.3@osu.edu.}
 \address[Second]{Department of Mechanical and Aerospace Engineering, The Ohio State University,   Columbus, OH 43212 USA (e-mail: paugh.29@buckeyemail.osu.edu, canova.1@osu.edu, stockar.1@osu.edu.)}

\begin{abstract}                

Connected and Automated Vehicles (CAVs) offer significant potential for improving energy efficiency and lowering vehicle emissions through eco-driving technologies. Control algorithms in CAVs leverage look-ahead route information and Vehicle-to-Everything (V2X) communication to optimize vehicle performance. However, existing eco-driving strategies often neglect macroscopic traffic effects, such as upstream traffic jams, that occur outside the optimization horizon but significantly impact vehicle energy efficiency. This work presents a novel Neural Network (NN)-based methodology to approximate the terminal cost within a model predictive control (MPC) problem framework, explicitly incorporating upstream traffic dynamics. By incorporating traffic jams into the optimization process, the proposed traffic-aware approach yields more energy-efficient speed trajectories compared to traffic-agnostic methods, with minimal impact on travel time. The framework is scalable for real-time implementation while effectively addressing uncertainties from dynamic traffic conditions and macroscopic traffic events.
\end{abstract}


\end{frontmatter}

\section{Introduction}
Eco-driving strategies for connected and automated vehicles (CAVs) have demonstrated significant potential to improve vehicle energy efficiency and reduce emissions by leveraging vehicle-to-everything (V2X) communication and onboard sensing technologies \citep{VAHIDI2018822, ozkan_SAE_2024}. These systems utilize real-time information to adjust vehicle speed and powertrain operations, optimizing energy consumption and travel time \citep{rajakumar2021vehicle, kavas2023comprehensive}. However, effectively deploying such strategies remains a complex challenge, largely due to the variability and uncertainty of traffic conditions \citep{LiReview2024}.

One common approach incorporates traffic considerations by modeling interactions with lead vehicles within the optimization horizon, as presented by \citep{Li2024Traffic}. This method accounts for traffic dynamics within the receding horizon by representing the lead vehicle using a stochastic vehicle model designed to capture the variability in upstream traffic. Similarly, other studies \citep{SU2022700, Ma2022} incorporate upstream traffic conditions in the eco-driving problem through a lead vehicle, which is represented by the regulatory driving cycles such as US06. While these methods can improve short-term decision-making in the eco-driving problem, it has inherent limitations. First, it requires the macroscopic traffic conditions simulated by the lead vehicle to be within the ego vehicle’s proximity, making it incapable of addressing upstream traffic events, such as a traffic jam occurring outside the range of vehicle detection \citep{ozkan2021eco}. 
Second, this approach does not differentiate between fundamentally different traffic scenarios, such as a slow-moving individual vehicle and a large-scale traffic jam, treating them equivalently despite their distinct impacts on energy efficiency \citep{de2015traffic}. Addressing these limitations requires a holistic framework that bridges macroscopic traffic events beyond the horizon with microscopic control strategies.  \par 
Recent studies have explored strategies to address these limitations by expanding the focus beyond direct interactions with immediate lead vehicles. For instance, studies such as \citep{Hu2023Generic, Lu2024} incorporate distinct traffic dynamics into their eco-driving framework, accounting for proximal disturbances caused by the lead vehicle and vehicle queueing at the intersections. These approaches differentiate these scenarios to enable more effective optimization of energy-efficient trajectories. However, these methods still depend on the presence of the upstream traffic conditions within the ego vehicle's proximity, limiting their ability to address any other traffic conditions beyond the optimization horizon.

This paper introduces a novel traffic-aware neural network-based model predictive control (MPC) framework for eco-driving control of CAVs. By explicitly incorporating macroscopic traffic dynamics beyond the optimization horizon into the terminal cost approximation, the framework bridges macroscopic traffic events with microscopic control strategies. This approach enables smoother and more energy-efficient speed trajectories in congested traffic scenarios.
\vspace{-1.5mm}
\section{Vehicle Dynamics \& Powertrain Model}
\vspace{-1.5mm} 
The vehicle considered in this paper is a Chrysler Pacific Plugin Hybrid Electric Vehicle (PHEV) presented by \citep{Pittel_SAE_2018} and consists of three power sources: a 3.6L engine and two permanent magnet synchronous AC electric machines i.e., Motor-Generator A (MGA) and Motor-Generator B (MGB) in a combined power-split configuration for a combined 260 HP. A calibrated model of the longitudinal vehicle dynamics and a power-split PHEV powertrain presented by \citep{ozkan_SAE_2024} is used in this work for computing energy use. The model was validated with test data collected on a closed course test track over various regulatory drive cycles.
\vspace{-1.5mm}
\section{Problem Formulation}
\vspace{-1.5mm} 
\cite{Paugh_ASME_DSCC_2023} presented a novel receding-horizon optimization framework for the eco-driving problem, which featured a Neural Network (\emph{NN})-based method for the terminal cost approximation. The powertrain used in that work was a 48V mild hybrid configuration. In this work, a similar methodology for NN-based terminal cost approximation has been extended to the PHEV powertrain. The NN-based approach has been extended further to account for variability due to traffic conditions in the terminal cost approximation. Specifically, two (2) NNs have a been trained i.e., a Traffic-Agnostic Neural Network (\emph{T-Ag-NN}) and a Traffic-Aware Neural Network (\emph{T-Aw-NN}). First, the formulation of the full-route optimization is presented, which is used to generate the dataset for training the \emph{T-Ag-NN}. This is then followed by formulation of the structure for the \emph{T-Aw-NN} and later a comparison of the results from the two NNs against a global optimal full-route solution.

The eco-driving problem is formulated in the distance domain to account for distance-based constraints. Let $s \in [0,N] \subset \mathbb{R}$ denote the discrete distance step, $x_{s} = [v_{s},\xi_{s}, t_{s}]^{\intercal} \in \mathcal{X} \subset \mathbb{R}^{n}$ the state variables comprising of vehicle velocity $v_{s}$, battery state of charge (SoC) $\xi_{s}$ and travel time $t_{s}$, $u_{s} = [a_{s}, \phi_{s}]^{\intercal} \in \mathcal{U} \subset \mathbb{R}^{m}$ the control input comprising of the vehicle acceleration $a_{s}$ and engine on/off control $\phi_{s}$ respectively. 

The discretized state dynamics are governed by (\ref{eq:state-dynamics}) with known initial states at the beginning of the trip.
\begin{equation}
    x_{s+1} = f(x_{s},u_{s}), s \in [0, N-1]
    \label{eq:state-dynamics}
\end{equation}
The equations governing the discrete state dynamics $f(x_{s},u_{s})$ at distance step $s$ are defined by:
\begin{subequations}\label{eq:dynamics}
    \begin{align}
        v^{2}_{s+1} = v^{2}_{s} + 2 \Delta s \cdot \left( \frac{F_{tr,s} - F_{road,s}(v_{s})}{M}\right) \label{eq:velocity} \\
        \xi_{s+1} = \xi_{s} - \frac{\Delta s}{\bar{v}_s C_{nom}} \cdot \frac{V_{oc} - \sqrt{V_{oc}^2 -4 R_0(\xi_{s}) P_{dmd,s}}}{2R_{0}(\xi_{s})} \label{eq:soc} \\
        t_{s+1} = 
        \begin{cases}
        t_{s} + t_{RG,s} & \quad s \in \mathcal{D}_{TL} \, \text{and} \, \bar{v}_s = 0 \\
        t_{s} + \frac{\Delta d_s}{\bar{v}_s} & \quad s \notin \mathcal{D}_{TL} 
        \end{cases} \label{eq:time}
    \end{align} 
\end{subequations}
where $F_{tr,s}$ and $F_{road,s}$ are the tractive force and road load resistive force respectively, $M$ the equivalent vehicle mass, $\bar{v}_s$ is the average velocity, $C_{nom}$ the nominal battery capacity, $V_{oc}(\xi_s)$ is the open circuit voltage, $R_{0}(\xi_s)$ the internal resistance, $P_{dmd,s}$ the electrical power demand, $t_{RG,s}$ the time remaining in the red phase of the traffic light at position $s$, with position of traffic lights contained in the set $\mathcal{D}_{TL}$ know from advanced mapping information.

The box constraints in (\ref{eq:constraints}) that can generally be expressed in the form $ h(x_{s},u_{s}) \leq 0$  are imposed:
\begin{equation}\label{eq:constraints}
\begin{aligned}
        x_{s}^{min} \leq x_{s} \leq x_{s}^{max} \\
        u_{s}^{min} \leq u_{s} \leq u_{s}^{max} \\
        t_s \in {\mathcal{T}}_{G,s}, \, t_s \geq \hat{t}_s^{l} + t_{\text{gap}} \\
\end{aligned}
\end{equation}
where ${\mathcal{T}}_{G,s}$ is the permissible green window for the traffic lights, $\hat{t}_s^{l}$ is predicted travel time of the lead vehicle using any of the lead-velocity predictor models presented by \citep{Lakshmanan_SAE_2023}, and $t_{\text{gap}}$ is a calibrated safety time gap of $2\mathrm{s}$ based on commercial ACC systems studies \citep{makridis2020TRR}.
\subsection{Full Route Optimization}
The full-route solution is used as the benchmark to compare the performance of the NN-based approaches against. To perform a full-route optimization that is used for training the \emph{T-Ag-NN}, it is assumed that the traffic light timing and phase information for all the traffic lights along the routes used in the training set is known, and that the traffic lights follow a fixed-pattern cycle. Actuated traffic lights are not considered in this work. To perform a full-route optimization that is used for training the \emph{T-Aw-NN}, in addition to signal phase and timing (SPaT) information, it is assumed that the lead-vehicle trajectories and traffic jam information for the entire route are known a-priori.  

With a sequence of admissible control policies $\mu: (\mu_s)_{s=0}^{N-1}$ that satisy the constraints in (\ref{eq:constraints}), the stage cost $c: \mathcal{X} \times \mathcal{U} \to \mathbb{R}$ to minimize at every step is defined by (\ref{eq:stage-cost}):
\begin{equation}\label{eq:stage-cost}
    \begin{aligned}
        c(x_{s}, \mu_{s}(x_s)) = \left( \gamma \frac{\dot{m}_{f,eq,s}(x_s, \mu_s)}{\dot{m}_{f}^{norm}} + (1-\gamma) \right) \cdot t_s
    \end{aligned}
\end{equation}
where $\dot{m}_{f,eq,s}(x_s, \mu_s) = \dot{m}_{f,s} + k_{batt} \frac{P_{batt}}{Q_{LHV}}$. $\gamma$ is the tradeoff factor between fuel efficiency and travel time, ranging from 0 to 1, $\dot{m}_{f,s}$ is the fuel flow rate from steady-state maps, $P_{batt}$ is battery power, $Q_{LHV}$ is the lower heating value of the fuel, and $k_{batt}$ is a conversion constant for converting battery power usage to an equivalent fuel flow rate. The full-route Optimal Control Problem (OCP) is formulated over $N$ steps as:
\begin{equation}\label{eq:fullroute-ocp}
\begin{aligned}
        \min_{{\lbrace{\mu}_s\rbrace}_{s=0}^{N-1}} c_{N}(x_{N}) +  \sum_{s=0}^{N-1} c(x_{s}, \mu_{s}(x_s)) \\
        \mathrm{s.t.} \quad x_{s+1} = f(x_{s},\mu_{s}(x_{s})); \quad h(x_{s},\mu_{s}(x_{s})) \leq 0
\end{aligned}
\end{equation}
where $c_{N}(x_{N})$ is the terminal cost. This full-route optimization in (\ref{eq:fullroute-ocp}) is solved using a deterministic Dynamic Programming (\emph{DP}) approach proposed by \citep{bertsekas1995dynamic}. This is done over various routes and their corresponding traffic light SPaT conditions, and the value functions are stored and used in the NN-training process as shown by \citep{Paugh_ASME_DSCC_2023} to train the \emph{T-Ag-NN}.
\vspace{-1mm}
\subsection{Receding Horizon Optimization}
For real-time implementation, the problem is formulated as an MPC framework, and solved using the Rollout algorithm, as was done by \citep{DeshpandeASME2022} with a goal of minimizing the cost functional shown in (\ref{eq:rhocp-formulation})
\begin{equation}\label{eq:rhocp-formulation}
    \begin{aligned}
        \mathcal{J}^{*}(x_s) = \min_{{\lbrace{\mu}_k\rbrace}_{k=s}^{s+N_H-1}} c_{T}(x_{s+N_H})+\sum_{k=s}^{s+N_H-1} c(x_k, \mu_{k}(x_k))\\
        \mathrm{s.t.} \quad x_{s+1} = f(x_{s},\mu_{s}(x_{s})); \quad h(x_{s},\mu_{s}(x_{s})) \leq 0
    \end{aligned}
\end{equation}

\begin{figure}[ht]
    \begin{center}
    \includegraphics[width=8.8cm]{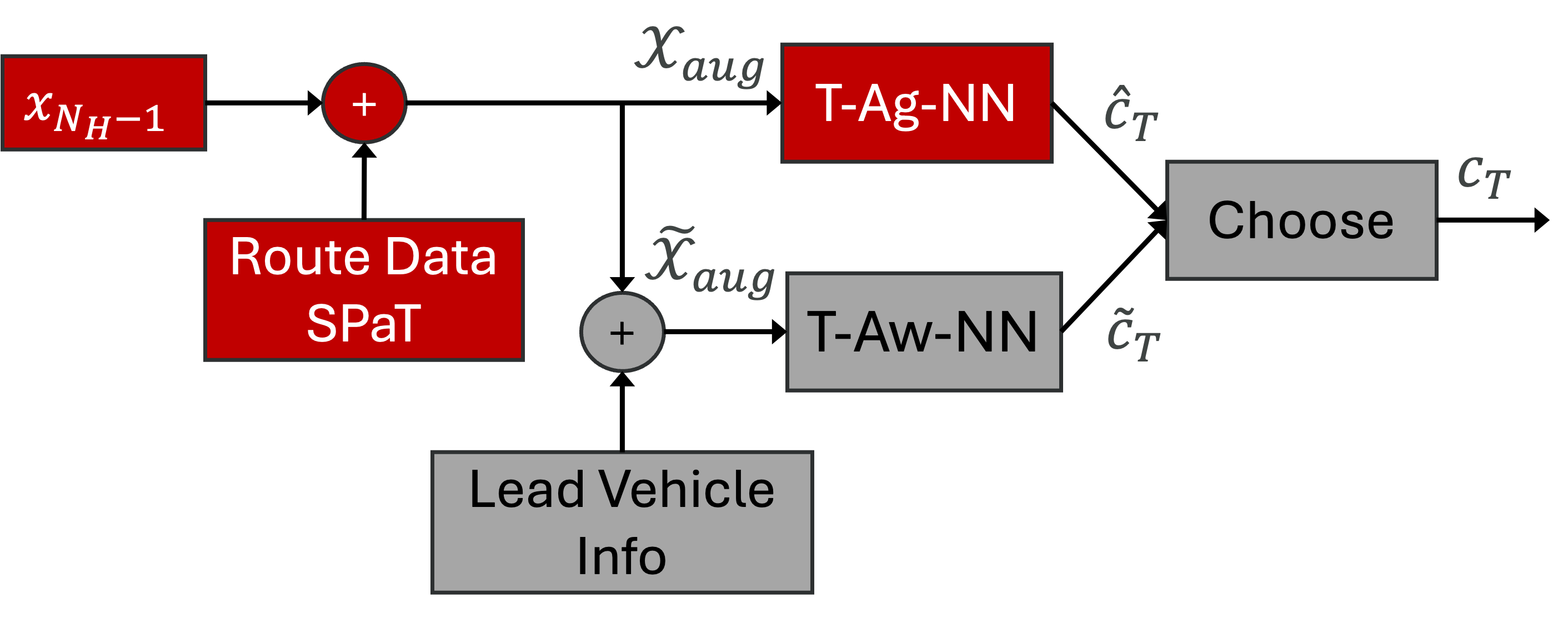}    
    \caption{Ensemble-NN (\emph{T-Ag-NN} \& \emph{T-Aw-NN}).} 
    \label{fig:nn-ensemble}
    \end{center}
\end{figure}

\cite{DeshpandeASME2022} used the value function from the full-route DP solution as the terminal cost $c_{T}(x_{s+N_H})$ at the end of each horizon. For this work, \emph{T-Ag-NN} and \emph{T-Aw-NN} generate different terminal cost approximations, i.e., $\hat{c}_{T}(x_{s+N_H}), \, \Tilde{c}_{T}(x_{s+N_H})$ respectively that are used in the MPC in (\ref{eq:rhocp-formulation}). For any of the MPCs that utilize different terminal cost approximations, the inclusion of traffic or traffic lights within the $N_{H}$ horizon is guaranteed by respecting the constraints in (\ref{eq:constraints}). The NNs are implemented as an ensemble shown in Fig. \ref{fig:nn-ensemble}, where if a lead vehicle is detected within the $N_{H}$ horizon, the \emph{T-Aw-NN} becomes active and generates a terminal cost $\tilde{c}_T(x_{s+N_H})$; otherwise, the system defaults to the \emph{T-Ag-NN} and generates a terminal cost $\hat{c}_T(x_{s+N_H})$. In these MPCs, distance step $\Delta s$ and horizon $N_{H}$ are set to 10 m and 200 m, respectively. Readers are referred to \citep{ozkan_SAE_2024} and \citep{ozkan2025modeling} for further details of the state and control input constraints of the applied MPC framework.
\section{Traffic Agnostic Neural Network (T-Ag-NN)}
\vspace{-1.5mm}
A fully connected feed-forward neural network is trained using the same process presented by \citep{Paugh_ASME_DSCC_2023}. The training dataset is generated from route information available via advanced mapping and value functions from full-route solutions over various route and SPaT conditions. The augmented state vector ${\mathcal{X}_{\text{aug}}}$ that serves as the input vector to the trained \emph{T-Ag-NN} is presented in Table \ref{tab:ag-nn-state-action-space} and using the \emph{Ensemble-NN}, the \emph{T-Ag-NN} generates $\hat{c}_T(x_{s+N_H})$ that is used in (\ref{eq:rhocp-formulation}) when lead vehicle information is not used to account for macroscopic effects in the terminal cost.
\begin{table}[H]
   \caption{Input Vector for \emph{T-Ag-NN}}
    \centering
    \begin{tabular}{l|c|l}
       & Variable & Description \\\hline
       & $\xi \in \mathbb{R} $  & Battery SoC \\ 
       & $v_{veh} \in  \mathbb{R}$ & Vehicle velocity \\
       & $v_{rlim} \in  \mathbb{R} $ & Vel. offset from current speed limit \\
       & $v_{rlim}^{'} \in \mathbb{R}$ & Vel. offset from the next speed limit \\ 
       ${\mathcal{X}_{\text{aug}}}$  & $d_{\text{tfc}} \in \mathbb{R}$ & Distance to upcoming traffic light \\
       & $d_{lim}^{'} \in \mathbb{R}$ & Distance to next speed limit change \\
       & $d_{rem} \in \mathbb{R}$ & Remaining trip distance \\  
       & ${x}_{\text{tfc}} \in {\lbrace -1,1 \rbrace}^{6}$ & \makecell[l]{Sampled status of upcoming traffic \\ light.} \\
\end{tabular}
    \vspace{1mm}
 
    \label{tab:ag-nn-state-action-space}
\end{table}

\section{Traffic-Aware Neural Network (T-Aw-NN)}
\vspace{-3mm}
To account for the uncertainty in driving conditions arising from macroscopic traffic events, \emph{T-Ag-NN} is extended by including traffic jam information. In the training process of the \emph{T-Aw-NN}, a traffic jam representation is introduced. A traffic jam is represented by a spatial-temporal variation of the route speed limit, effectively resulting in a variation in traffic density $\rho_{tfc}$. The correlation between the speed limit of the route and $\rho_{tfc}$ is defined by Greenshield's equation (\ref{eq:greenshield}) where $v_{jam}$ is the average speed of traffic in a traffic jam, and $c_1, c_2$ are model constants \citep{rakha2002comparison}.
\begin{equation}\label{eq:greenshield}
    \begin{aligned}
        v_{jam} = c_2 - c_1 \cdot \rho_{tfc}
    \end{aligned}
\end{equation}

The speed limit is subject to an additional traffic jam constraint as defined in (\ref{eq:jam-effect-on-speed})
\begin{equation}\label{eq:jam-effect-on-speed}
    \begin{aligned}
        v^{lim}(x_s,t_s) = 
        \begin{cases}
            v_{jam}, \quad \underline{t}_{s} < \bar{t}_s, \underline{x}_{s} < \bar{x}_s \\
            v_{s}^{max}, \quad \mathrm{otherwise}
        \end{cases}
    \end{aligned}
\end{equation}
where $\underline{t}_{s}, \bar{t}_{s}$ denote the start and end time of the jam respectively, and $\underline{x}_{s}, \bar{x}_{s}$ denote the start and end position of the jam respectively. The time-varying traffic jam can then be implemented in the full-route optimization in \ref{eq:fullroute-ocp} as a velocity constraint by including (\ref{eq:velocity-constraint})
\begin{equation}\label{eq:velocity-constraint}
    \begin{aligned}
        v_{s}(x_s, t_s) \in [v_{s}^{min}, v^{lim}(x_s,t_s)]
    \end{aligned}
\end{equation}

The \emph{T-Aw-NN} is trained by additionally including information about the lead vehicle and/or traffic jam in the training process. The additional inputs to the augmented input vector for training \emph{T-Aw-NN} are presented in Table \ref{tab:aw-nn-state-action-space} and using the \emph{Ensemble-NN}, the \emph{T-Aw-NN} generates $\Tilde{c}_T(x_{s+N_H})$ that is used in (\ref{eq:rhocp-formulation}) when lead vehicle information is used to account for macroscopic effects in the terminal cost.

\begin{table}[H]
\caption{Additional Input Vector for \emph{T-Aw-NN}}
    \centering
    \begin{tabular}{c|l|l}
       & Variable & Description \\\hline
       & $d_{k}^{l} \in \mathbb{R} $  & Distance to lead vehicle \\ 
       $ \mathcal{X} \in \Tilde{\mathcal{X}}_{aug}$  & $v_{k}^{l} \in  \mathbb{R}$ & Relative velocity to lead vehicle \\
       $\mathcal{X} \notin {\mathcal{X}_{aug}}$ & $t_{k}^{l} \in  \mathbb{R} $ & Time to reach lead vehicle \\
\end{tabular}
    \vspace{1mm}
    
    \label{tab:aw-nn-state-action-space}
\end{table}

\begin{figure}[H]
    \begin{center}
    \includegraphics[width=8.8cm]{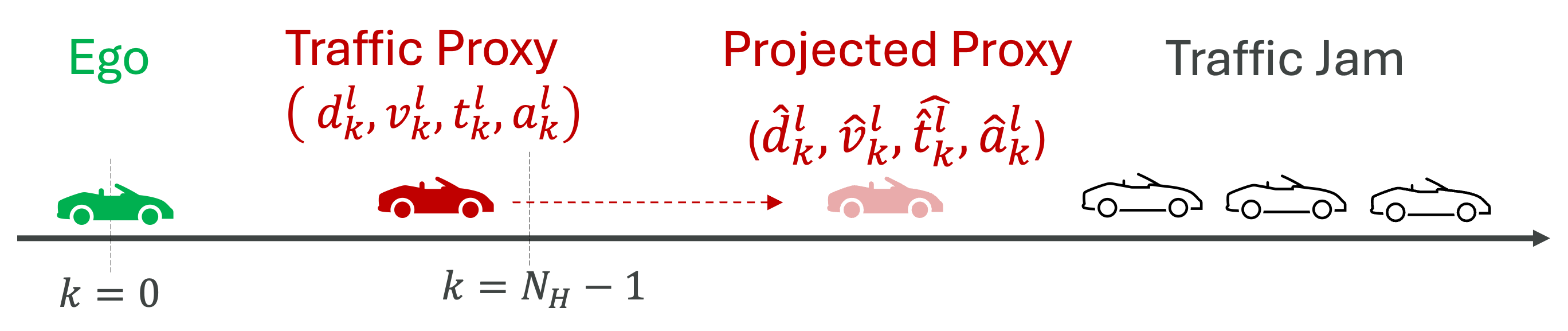}    
    \caption{Interaction of ego vehicle, traffic proxy and upcoming traffic jam.} 
    \label{fig:traffic-proxy}
    \end{center}
\end{figure}

\begin{table*}
\centering
\caption{Performance comparison of \emph{T-Ag-NN} and \emph{Ensemble-NN} MPC strategies against the DP solution across equivalent fuel consumption, travel time, and final SoC for two routes.}
\begin{tabular}{@{}lccccccccc@{}}
\toprule
\toprule
  & \multicolumn{3}{c}{Equivalent Fuel Consumption [g]} & \multicolumn{3}{c}{Travel Time (s)} & \multicolumn{3}{c}{Final SoC (\%)} \\ 
\cmidrule(lr){2-4} \cmidrule(lr){5-7} \cmidrule(lr){8-10}
      & DP & T-Ag-NN & Ensemble-NN & DP & T-Ag-NN & Ensemble-NN & DP & T-Ag-NN & Ensemble-NN \\ \midrule
Route 1   & 25.6 & 72.3 & 67.6 (-6.5\%) & 176.8  & 183.9 & 187.1 (+1.7\%) & 24.2 & 25.0 & 25.2 \\
Route 2  & 175.2 & 200.7 & 195.4 (-2.6\%) & 490.6  & 488.6 & 488.7 (+0.1\%) & 25.7 & 25.2 & 25.2  \\
\bottomrule
\bottomrule
\end{tabular}
    \vspace{.1mm}
\label{tab:summary_statistics}
\end{table*}
For online implementation in the MPC, the parameters used for the training process specifically $v_{jam}, \underline{t}_{s}, \bar{t}_{s},\underline{x}_{s}, \bar{x}_s$ are unknown. The proposed solution is to use a traffic proxy as shown in Fig. \ref{fig:traffic-proxy}, which interacts with the ego vehicle within the $N_{H}$ horizon. The behavior of the proxy is projected for the next $200\mathrm{m}$ beyond the horizon using a constant acceleration model \citep{Lakshmanan_SAE_2023}, and can interact with a traffic jam. The additional inputs to the augmented input vector for \emph{T-Aw-NN} are extracted from this projection of the traffic proxy vehicle, and this process generates the $\hat{c}_T(x_{s+N_H})$ used in (\ref{eq:rhocp-formulation}).

This interaction between the ego vehicle and the traffic proxy shown in Fig. \ref{fig:traffic-proxy} with the projection outside the $N_H$ horizon allows the traffic proxy's extrapolated behavior to inform the running cost in (\ref{eq:rhocp-formulation}), while the traffic proxy's extrapolated behavior beyond the horizon, where it can interact with traffic jams, informs the terminal cost $\hat{c}_T(x_{s+N_H})$ which includes macroscopic traffic effects outside the ego vehicle's horizon. The hyper-parameters used for training the Ensemble-NN are presented in \citep{paugh:thesis:2023}.
\vspace{-2mm}
\section{Results and Discussion}
\vspace{-2mm}
\subsection{Simulation Setup}
\vspace{-1.5mm}
The simulation study evaluates the performance of the proposed \emph{Ensemble-NN} MPC, \emph{T-Ag-NN} MPC, and benchmark DP eco-driving strategies on two representative routes, each route designed to reflect urban driving scenarios involving signalized intersections and downstream congestion:
\begin{itemize}
    \item \textbf{Route 1:} A 2 km route containing two traffic lights and a traffic jam located near the end of the route.
    \item \textbf{Route 2:} A 5 km route with four traffic lights and a longer downstream congestion zone.
\end{itemize}
In both cases, the traffic jam is modeled as a low-speed region where the lead vehicle travels at a constant 10 m/s. The length of the jam varies between routes, creating different levels of interaction between the ego and lead vehicles. These traffic jam regions are shown by shaded rectangles in Fig. \ref{fig:jam_1} and Fig. \ref{fig:jam_2}.\par 
To simulate the traffic proxy for the ego vehicle for both cases, lead vehicle trajectories are generated by using DP formulation in \eqref{eq:fullroute-ocp}, which serves as a proxy for realistic traffic behavior. It is worth mentioning that the proposed framework does not rely on this specific choice; the lead vehicle trajectory can also be defined by using driver models, alternative eco-driving methods, or real-world driving data.


\subsection{Simulation Results}
\vspace{-1.5mm}
Table~\ref{tab:summary_statistics} compares the performance of the \emph{T-Ag-NN} and proposed \emph{Ensemble-NN} MPC strategies against the wait-and-see DP benchmark across three metrics: equivalent fuel consumption (EFC), travel time, and final battery SoC for two simulated routes. The wait-and-see DP solution, which assumes perfect future traffic knowledge, achieves the lowest EFC on both routes. However, this optimal performance is based on idealized assumptions and is not feasible for real-time deployment. Among the two NN-based MPC strategies, the \emph{Ensemble-NN} MPC consistently outperforms the \emph{T-Ag-NN} MPC in terms of energy efficiency while maintaining comparable travel time and final SoC. In Route 1, the \emph{Ensemble-NN} MPC achieves a 6.4\% improvement over the \emph{T-Ag-NN} MPC, with only a 1.7\% increase in travel time. In Route 2, the \emph{Ensemble-NN} MPC achieves a 2.6\% reduction in EFC compared to \emph{T-Ag-NN}  MPC, with minimal difference in travel time between the two approaches.

\begin{figure}[H]
    \centering
    \subfigure{{\includegraphics[angle=0, scale=0.20]{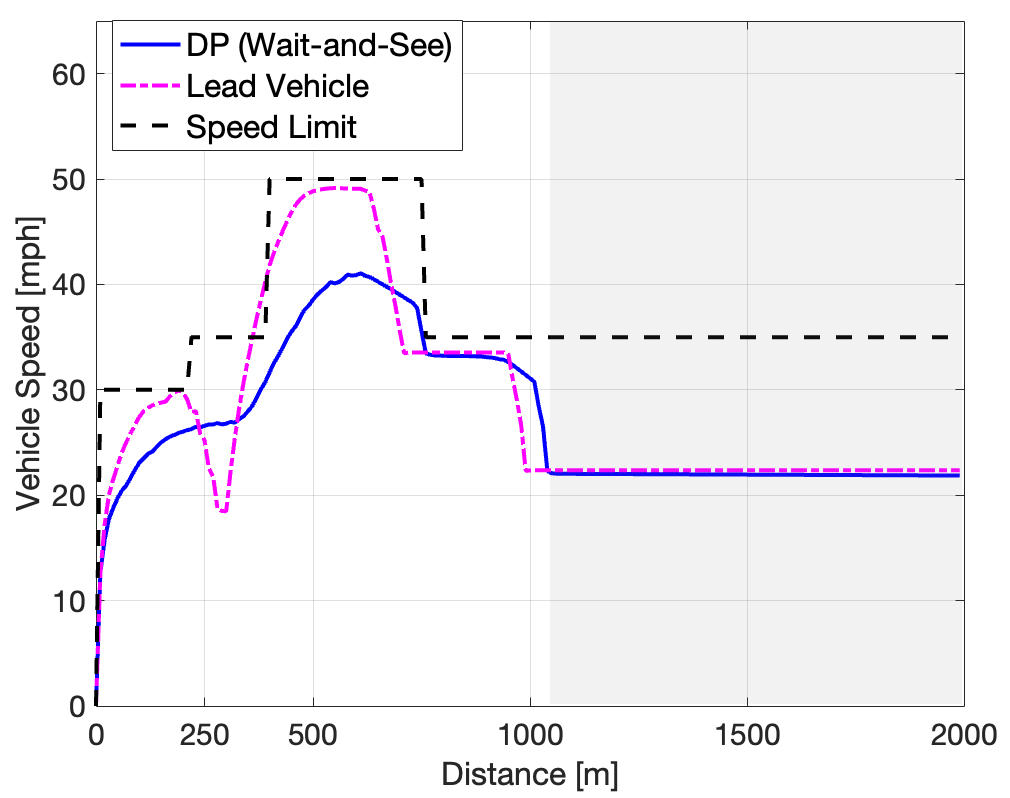}}}
    \vspace{-4mm}
    \subfigure{{\includegraphics[angle=0, scale=0.20]{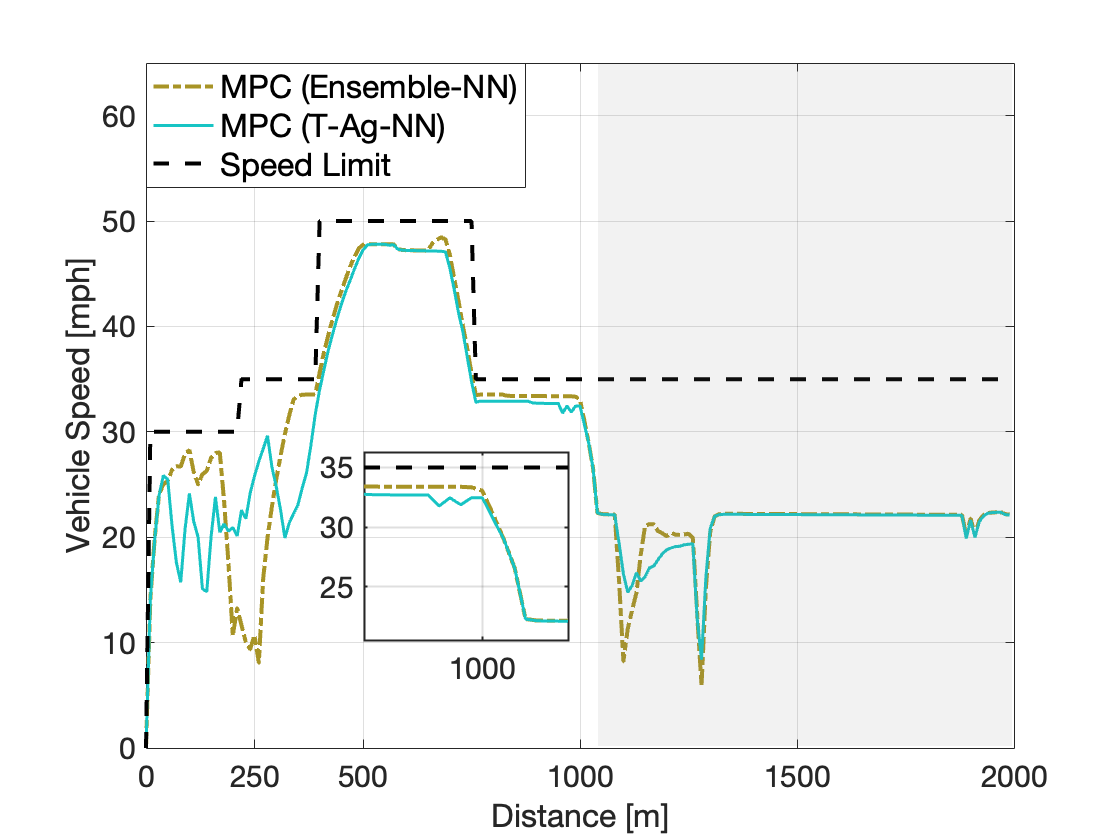}}}
    \vspace{-4mm}
    \subfigure{{\includegraphics[angle=0, scale=0.20]{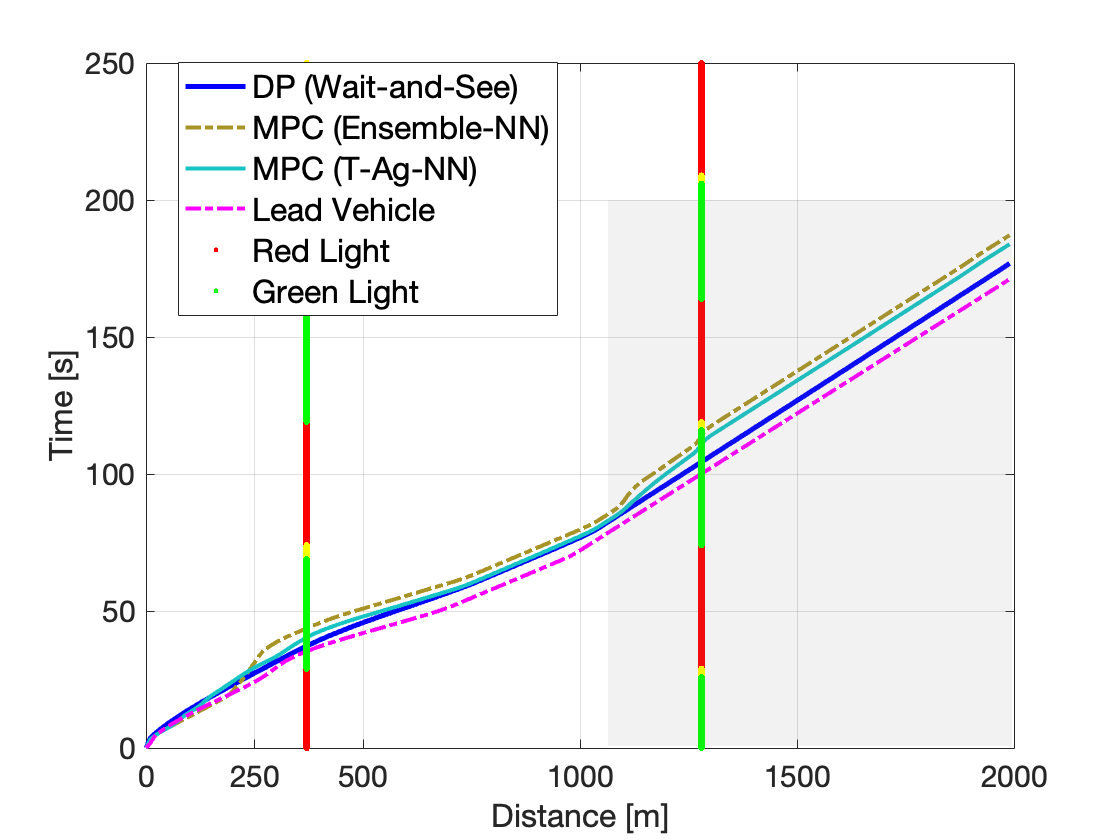}}}
    \vspace{-4mm}
    \subfigure{{\includegraphics[angle=0, scale=0.20]{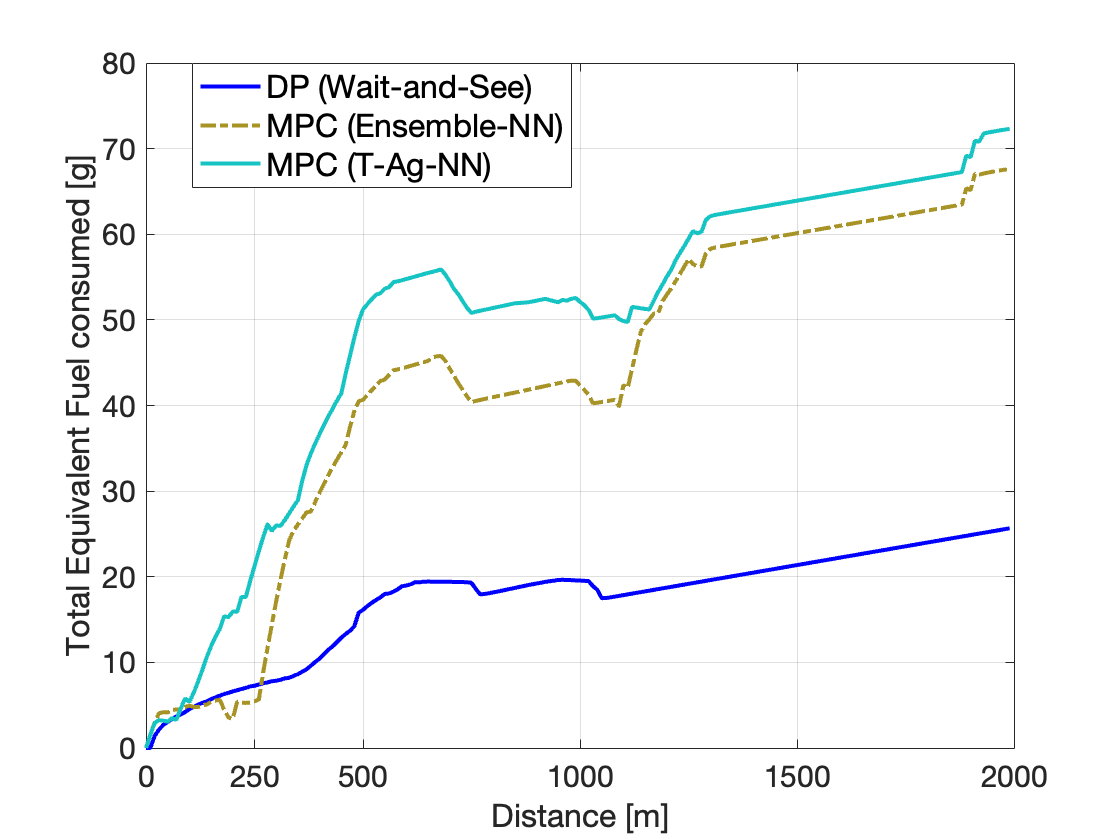}}}
          \caption{Comparison of wait-and-see DP, Ensemble-NN and T-Ag-NN MPC solutions in Route 1.}
    \label{fig:jam_1}
\end{figure}

\begin{figure}[H]
    \centering
    \vspace{-3.25mm}
    \subfigure{{\includegraphics[angle=0, scale=0.20]{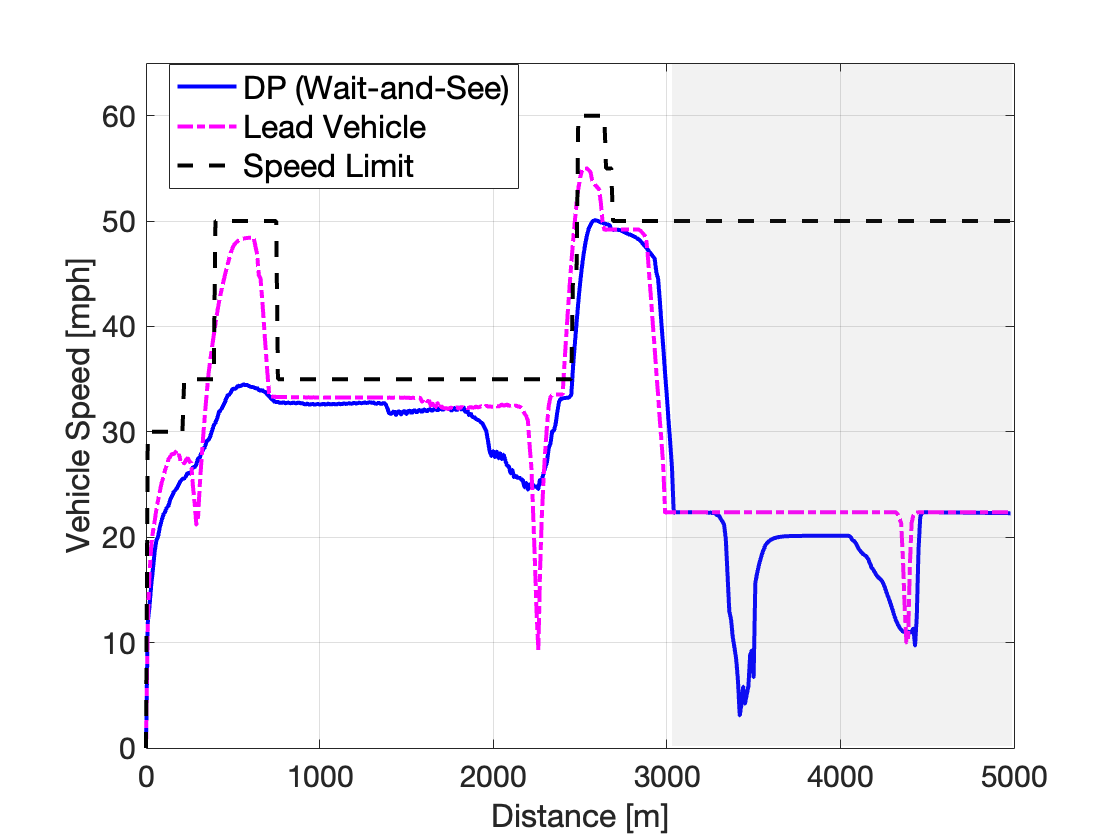}}}
    \vspace{-4mm}
    \subfigure{{\includegraphics[angle=0, scale=0.20]{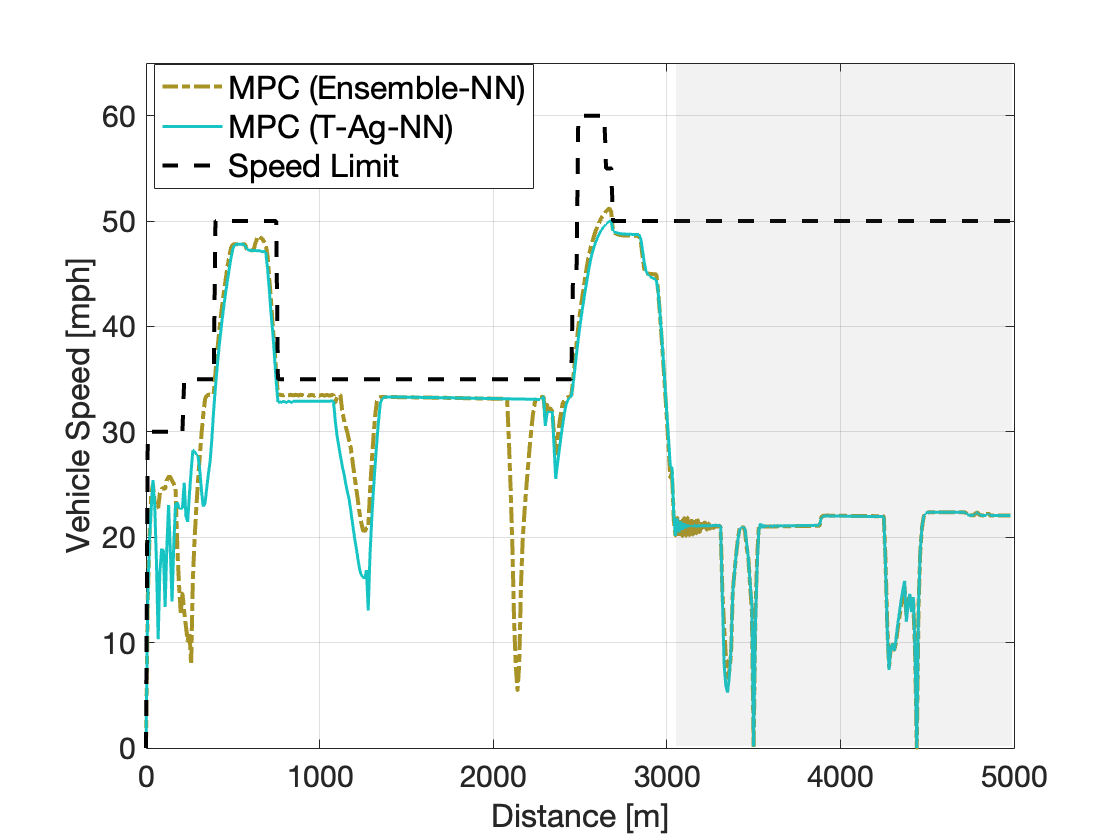}}}
    \vspace{-4mm}
    \subfigure{{\includegraphics[angle=0, scale=0.20]{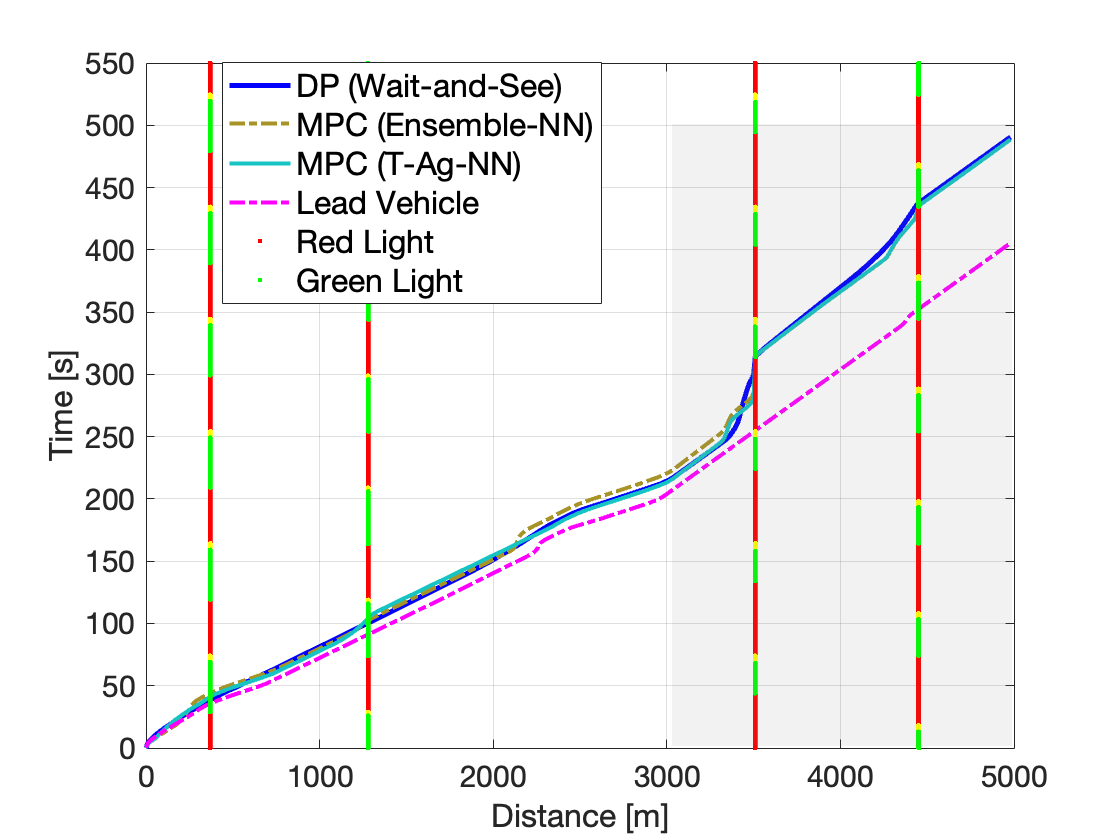}}}
    \vspace{-4mm}
    \subfigure{{\includegraphics[angle=0, scale=0.20]{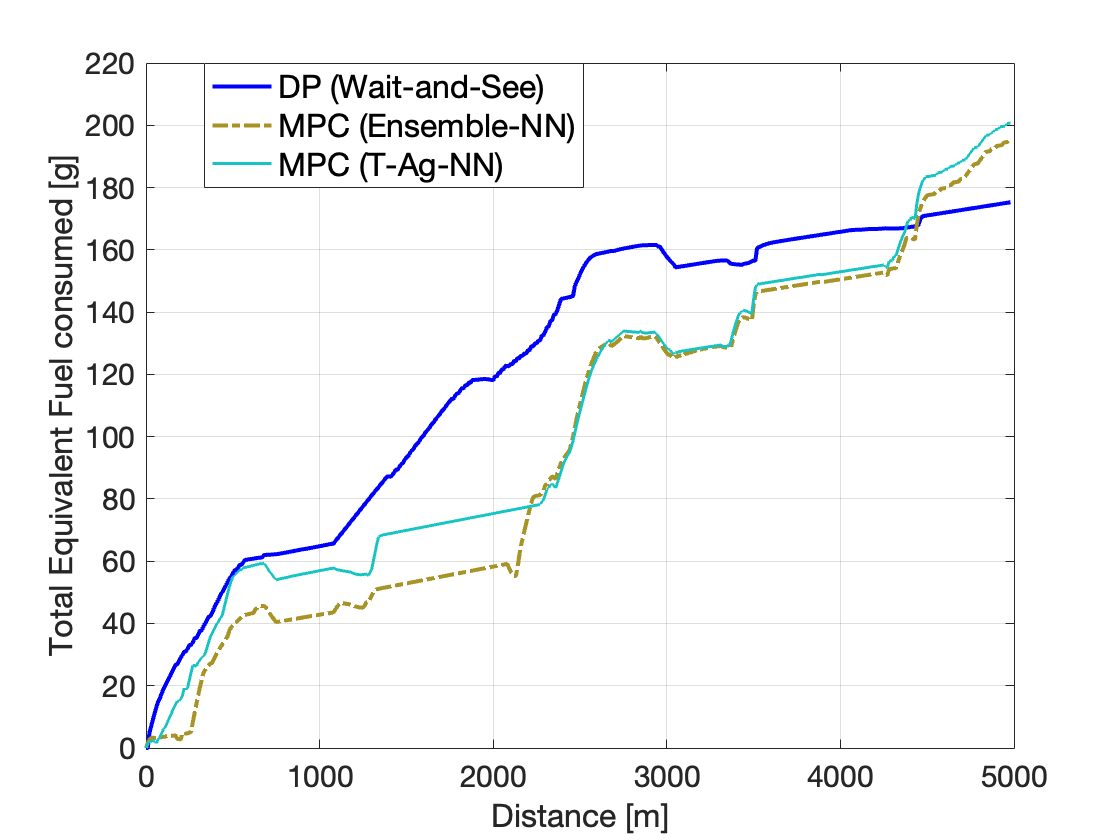}}}

    \caption{Comparison of wait-and-see DP, Ensemble-NN and T-Ag-NN MPC solutions in Route 2.}
    \label{fig:jam_2}
\end{figure}

Figures~\ref{fig:jam_1} and~\ref{fig:jam_2} illustrate how the proposed \emph{Ensemble-NN} MPC yields more energy-efficient trajectories than \emph{T-Ag-NN} MPC, particularly within the first 500 meters of both routes, where the ego vehicle closely follows the lead vehicle. This behavior highlights the \emph{Ensemble-NN} MPC’s ability to incorporate projected lead vehicle behavior beyond the optimization horizon (200 m), allowing it to manage microscopic control more effectively. Additionally, \emph{Ensemble-NN} MPC demonstrates a more proactive braking strategy. This is especially noticeable around the 200-m mark in both routes, where \emph{Ensemble-NN} MPC anticipates the lead vehicle’s braking due to an upcoming traffic light and adjusts its speed with an early braking maneuver. In contrast, the \emph{T-Ag-NN} MPC responds with a more delayed and sudden braking action, resulting in less smooth and less energy-efficient driving behavior.

Further differences are observed when the ego vehicle approaches traffic congestion. For example, in Route 1 between 950 and 1050 meters, a closer examination in Figure~\ref{fig:jam_1} reveals that the \emph{Ensemble-NN} MPC adapts to the reduced traffic speed more smoothly than the \emph{T-Ag-NN} MPC. This zoomed-in view highlights \emph{Ensemble-NN}’s capacity to proactively adjust its speed in response to anticipated traffic patterns, thereby avoiding abrupt speed changes and maintaining a more stable trajectory. In contrast, the \emph{T-Ag-NN} MPC exhibits more reactive behavior with speed fluctuations.

By leveraging predictions of lead vehicle behavior and upcoming traffic changes beyond the immediate horizon, the \emph{Ensemble-NN} MPC achieves smoother and more energy-efficient driving behavior. These results indicate that the proposed approach can offer notable energy savings with minimal impact on travel time, demonstrating its effectiveness for real-time applications where traffic awareness is crucial.

\section{Conclusion}
This paper presented a traffic-aware neural network-based MPC framework for eco-driving control of CAVs. Traditional eco-driving methods typically focus on short-term optimization within a constrained horizon, limiting their ability to anticipate macroscopic traffic events, such as traffic jams or the behavior of lead vehicle beyond the horizon. The proposed framework addresses this limitation by integrating anticipated macroscopic traffic dynamics into terminal cost approximation within a hierarchical multi-horizon optimization framework.

Simulation results across varied route conditions demonstrated that the proposed framework achieves significant energy savings compared to a traffic-agnostic counterpart, with improvements in energy efficiency of up to 6.5\%. These results underscore the importance of incorporating macroscopic traffic dynamics into eco-driving optimization and highlight the framework’s potential to enhance real-time vehicle control in dynamic and stochastic traffic environments.

Future work will focus on extending the framework to handle more complex traffic scenarios, such as interactions with multiple traffic jams. Additionally, integrating real-world traffic data, such as GPS or infrastructure-based sensors, will help evaluate the framework's scalability and robustness under diverse driving conditions, key for practical implementation in urban environments.
\begin{ack}
\vspace{-3mm} 
The authors acknowledge the support from the United States Department of Energy, Advanced Research
Projects Agency-Energy (ARPA-E) NEXTCAR Program (Award Number DE-AR0000794) that made this work possible.
\end{ack}

\bibliography{ AAC2025_V2.bib}
\end{document}